\documentclass[referee]{aa}
\usepackage{graphicx}
\begin{document}


  \title{Collisional depolarization and transfer rates of spectral lines 
by atomic hydrogen. IV: application to ionised atoms.}

   \author{M. Derouich
          \inst{1},
           S. Sahal-Br\'echot
          \inst{1},
             and
          P. S. Barklem
          \inst{2}
          }
\titlerunning{Collisional depolarization  and transfer rates.}
\authorrunning{M. Derouich et al}
   \institute{Observatoire de Paris-Meudon, LERMA UMR CNRS 8112, 5, Place Jules Janssen, F-92195 Meudon Cedex, France. 
          \and
Department of Astronomy and Space Physics, Uppsala University, Box 515, S 
751 20 Uppsala, Sweden\\
   \email{Moncef.Derouich@obspm.fr}
             }
   \date{Received 2004 / accepted XXXX}

\abstract{
The semi-classical theory of collisional depolarization of spectral lines of
neutral atoms by atomic hydrogen (Derouich et al. \cite{Derouich_03a}; Derouich et al. \cite{Derouich_03b}; Derouich et al. \cite{Derouich_04} and references therein) is extended to spectral lines 
of singly ionised atoms. In this paper we apply our general method to the 
particular cases of the $3d$  $^2D$ and 
 $4p$ $^2P$ states of the CaII ion and to the $5p$ $^2P$ state of the SrII 
ion. Analytical 
expressions of all rates as a function of local temperature are given. 
 Our results for the CaII ion are compared to recent quantum chemistry  
calculations. A discussion of our results is presented. 
}
\maketitle
\section{Introduction}
Observations of the linearly polarized  radiation at the limb of the Sun 
(known as the ``second solar spectrum''), which is formed 
by coherent scattering processes, show rich structures 
(Stenflo \& Keller \cite{Stenflo_97}; Stenflo et al. \cite{Stenflo_00}; Stenflo  \cite{Stenflo_01}; Gandorfer \cite{Gandorfer_00};  Gandorfer 
\cite{Gandorfer_02}). The linear polarization observations reported  in the atlas recently published by Gandorfer (\cite{Gandorfer_00,Gandorfer_02}) show significant polarization peaks in many spectral lines of 
ions, e.g. NdII  5249 \AA, EuII  4129 \AA, CeII  4062 \AA,  CeII  4083 \AA, 
Ba II D2  4554 \AA, ZrII 5350 \AA, etc.. Several surveys of the 
scattering polarization throughout the solar spectrum
(Stenflo et al. \cite{Stenflo_80,Stenflo_83a,Stenflo_83b}; see also the Q/I 
observations of Stenflo et al. \cite{Stenflo_00} and the full Stokes-vector 
observations of Dittmann et al. \cite{Dittmann_01}) have shown that 
ionised  lines such as SrII 4078 \AA $\;$ and the IR triplet of  CaII 
are two of the more strongly polarized.  The  interpretation of these 
observations leads to the indirect determination of the turbulent magnetic 
field strength via the Hanl\'e effect. Such  interpretation requires the solution 
of the coupling between the polarized radiative transfer equations (RTE)
and the statistical equilibrium equations (SEE) taking into account the 
contributions of isotropic depolarizing collisions with neutral hydrogen. Depolarization and polarization transfer rates are currently available for ionised calcium levels, which have been obtained through sophisticated quantum  chemistry methods which are accurate but cumbersome. Indeed, it is very difficult and sometimes not  accurate to treat collision processes, involving heavy
ionised atoms like  Ti II, Ce II, Fe II, Cr II, Ba II..., by standard quantum chemistry 
methods.  
It would be 
useful to develop alternative methods capable of giving  
results for many levels of ionised atoms rapidly and with reasonable accuracy. 

The aim of this paper is 
to extend the semi-classical theory of collisional depolarization of spectral 
lines of neutral atoms by atomic hydrogen given in previous papers of this 
series (Derouich et al. \cite{Derouich_03a}; Derouich et al. \cite{Derouich_03b}; Derouich et al. \cite{Derouich_04}; 
hereafter Papers I, II and III respectively) to spectral lines of ions. This paper outlines the  
necessary adjustments to the theory presented in Papers I, II and III  for 
extension to  spectral lines of ions. Our results are  presented and 
compared  with those obtained in the case of CaII levels by the quantum chemistry 
method (Kerkeni et al. \cite{Kerkeni_03}). We also compare the results to the
depolarization rates computed with the Van der Waals potential. An advantange of the present method is that it is not specific for a given perturbed ion, and may be easily applied to any singly ionised species. 
Indeed, we have applied our method to calculate depolarization and polarization 
transfer rates for the upper state 5$p$ $^2P$ of the SrII 4078 \AA $\;$ line. 

The main feature of the technique is the use of perturbation theory in calculating the interatomic potentials.  A key parameter in this theory is $E_p$ which approximates the energy denominator in the second-order interaction terms by an average value (Paper I and ABO papers: Anstee \cite{Anstee2}; Anstee \& O'Mara \cite{Anstee1}, \cite{Anstee3}; Barklem \cite{Barklem3}; Barklem \& O'Mara \cite{Barklem1}; Barklem, O'Mara \& Ross \cite{Barklem2}).
A discussion of the effect of $E_p$ variation on depolarization rates is presented.  Finally, we show that the present semi-classical method
gives results in agreement with accurate but time consuming quantum chemistry calculations
to better than 15 $\%$ for the CaII ion ($T$= 5000K).  Using our method it should now be possible to rapidly 
obtain the data needed to interpret quantitatively the Stokes
parameters of the observed lines. 
\section{Statement of the problem}
Under typical conditions of formation of observed lines in the solar atmosphere,  the  atomic system (atom, ion or molecule) suffers isotropic collisions  with hydrogen atoms of the medium before it radiates. The states of the bath of hydrogen atoms are unperturbed.
 In the tensorial formulation (Fano \& Racah \cite{Fano1}; Messiah \cite{Messiah}; Fano \cite{Fano2}), the internal states of the perturbed 
particles (here these particles are singly ionised atoms)  are described by
the spherical tensor components $\displaystyle ^{nl J}\rho_q^k$ of 
the density matrix. Owing to the isotropy of the depolarizing collisions,  the depolarization rates,  polarization and population transfer rates are $q$-independent. The 
term corresponding to the depolarizing collisions in the master equation is 
given by
\begin{eqnarray} \label{eq1}
\big(\frac{d \; ^{n l J}\rho_0^{k}}{dt})_{coll} & = & - D^k(n l J, \; T) \; ^{n l J}\rho_0^k \nonumber \\
&+& \sum_{J \ne J'}  
D^k(n l J \to n l J', \; T) \;  ^{n l J'}\rho_0^k  \\  
&+&\textrm{quenching term.} \nonumber
\end{eqnarray}
$D^k(n l J, \; T)$ is the collisional depolarization rate of the ionic level 
$(nlJ)$ at the local temperature $T$ $(0 \; \le \; k \; \le \; 2J)$. $D^0(n l J, \; T)$ is the  destruction rate of population  which is zero since elastic collisions ($J=J'$) do not alter the population of the level $(nlJ)$. $D^1(n l J, \; T)$ is the  destruction rate of orientation (related to circular polarization) and $D^2(n l J, \; T)$ is the  destruction rate of alignment of the level $(nlJ)$ which is of interest in our astrophysics framework because it is related to the observed linear polarization.

$D^k(n  l J \to n  l J', \; T)$ is the polarization  transfer rate  between the levels $| nl J \rangle \to |nl J'  \rangle$,  where $0 \; \le \; k \; \le \;k_{\textrm{max}}$,  $k_{\textrm{max}}=$ 2$J$ if  $J<J'$ ( or if $J > J'$ then $k_{\textrm{max}}=$ 2$J'$). In particular,  $D^k(n  l J \to n  l J', \; T)$ corresponds to collisional  transfer of population $(k=0)$, orientation $(k=1)$ and alignment $(k=2)$.  

Higher order terms of $D^k(n l J, \; T)$ and $D^k(n l J \to n l J', \; T)$ with $k \; > \; 2$ can play a role in the SEE and have to be calculated.  Note that, for the  analysis of the linear polarization spectrum, only depolarization and polarization transfer rates with even $k$ are need. Odd $k$-terms can be eliminated from the SEE.

$D^k(n l J, \; T)$ and $D^k(n l J \to n l J', \; T)$ can be written as a linear combination of the collisional transition rates between the fine stucture sublevels $\zeta (n l J M_J \to n l J' M'_J, \; T)$ (Papers I, II and III, Sahal-Br\'echot \cite{Sahal2}); for depolarization rates $D^k(n l J, \; T)$ and transfer rate of population $D^0(n  l J \to n  l J', \; T)$, the coefficients of this linear combination are  positive while the signs of the coefficients of the linear combination  for transfer rates of rank $k\geq1$ may be either positive or negative. This explains why transfer rates of rank $k\geq1$ are significantly smaller  (Paper III). In our semi-classical theory,  the collisional transition rate between the sublevels $| nl J M_J \rangle \to |nl J  M'_J \rangle$ is given by (Paper I; Paper II):
\begin{eqnarray} \label{eq2}
\zeta(nl J M_J \to nl J M'_J, \; T) = n_H \int_{0}^{\infty} \int_{0}^{\infty} 2\pi b\, db\: v\; f(v)\; dv\: \nonumber \\
\times |\langle nl J M_J| I - S(b, v) |nl J M'_J \rangle|^2,
\end{eqnarray} 
where $f(v)$ is the Maxwell distribution of 
velocities for the local temperature $T$ and $n_H$ is the local hydrogen 
atom number density.
 $I$ is the unit matrix and $T = I-S $ is the so-called 
transition matrix depending on the impact-parameter $b$ and relative 
velocity $v$. The collisional depolarization rates and the 
collisional transfer rates, which are linear combinations of the $\zeta(nl J M_J \to nl J M'_J, \; T)$ given by equation (\ref{eq2}), can 
be expressed in terms of the $T$-matrix elements. The transition matrix $T$ is functionally dependent on the interaction energy matrix of hydrogen in its ground state  with the perturbed ion. Indeed, the transition
matrix elements in the dyadic basis are obtained by solving the time-dependent Schr\"odinger equation (Paper I)
\begin{eqnarray} \label{eq3}
(H_A+H_P+V_{\textrm{eff}}(R)) \big | \psi(t) \rangle =\textrm{i} \frac{d\big | \psi(t)\rangle}{dt}.
\end{eqnarray}
$V_{\textrm{eff}}$ is   the ion-hydrogen interaction used in this work and $| \psi(t) \rangle$ is the wave function  of the system (ion+hydrogen). $H_A+H_P$ is the Hamiltonian of the system at the interatomic distance $R=\infty$ (Fig. \ref{perturbed}). 
\section{Ion-hydrogen interaction potentials}
\begin{figure}[htbp] 
\begin{center}
\includegraphics[width=8 cm]{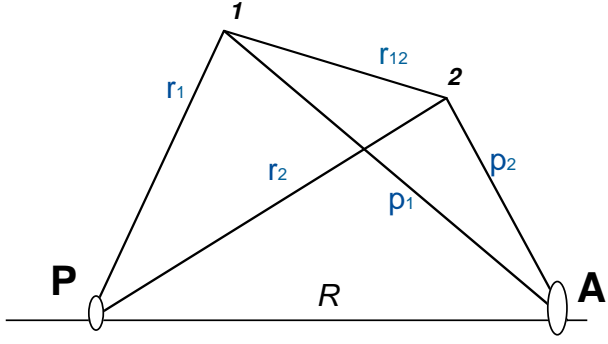}
\end{center}
\caption{The perturbed ion core (with charge $Z$=2) is located at A and the hydrogen perturbing 
core (a proton) at P. Their valence electrons are denoted by 1 and 2 
respectively.}
\label{perturbed}
\end{figure}

The interaction potential for a singly ionised atom interacting with a hydrogen 
is treated in much the same way as for the neutral atom interaction with
hydrogen  (Papers I, II and III; ABO papers). In the coordinate system of Fig. \ref{perturbed}, $V$ is  given, in atomic units, by (Barklem \& O'Mara \cite{Barklem3}):
\begin{equation} \label{eq4}
V=\frac{2}{R}+\frac{1}{r_{12}}-\frac{1}{r_2}-\frac{2}{p_1}= \frac{1}{R}+\frac{1}{r_{12}}-\frac{1}{r_2}-\frac{1}{p_1}+V_{\textrm{ind}}
\end{equation} 
and atomic units are used hereafter. $V_{\textrm{ind}}= 1/R -1/p_1$ is 
the part representing an inductive interaction between the excess charge
on the ionised atom and hydrogen atom. Quenching is neglected and thus we consider only the subspace $nl$ (2$l$+1 
states) and we denote the 
product state of the two separated atoms at $R=\infty$  by $|M_l\rangle$. By application of time-independent perturbation theory to
second-order, the interaction potential matrix elements are
given by:
\begin{eqnarray} \label{eq5}
\langle M_l | V_{\textrm{eff}} | M_l \rangle &=& \langle M_l | V |M_l\rangle \nonumber \\  
&+& \sum_{M'_l\ne M_l} \frac{ \langle M_l | V_{\textrm{ind}} | M'_l \rangle  
\langle M'_l |V_{\textrm{ind}} | M_l \rangle }{E_{M_l} - E_{M'_l}}. \nonumber \\
&+&\sum_{M'_l\ne M_l} 
\frac{ \langle M_l | V-V_{\textrm{ind}} | M'_l \rangle  
\langle M'_l |V-V_{\textrm{ind}} | M_l \rangle }{E_{M_l} - E_{M'_l}},   
\end{eqnarray} 
$ E_{M_l}$ are the unperturbed energy eigenvalues of the isolated atoms. The expression for the second-order interaction can be greatly simplified 
if we replace   the energy denominator $E_{M_l} - E_{M'_l}$, of each sum, 
 by a fixed average energy $E_p$ and assume that for important separations $E_p(R)=E_p(\infty)$.  This is  
the Uns\"old approximation (Uns\"old \cite{Unsold1}; Uns\"old \cite{Unsold2}). 
$E_p=-4/9$ atomic units is the appropriate Uns\"old energy value of the  part of  interaction, $V_{\textrm{ind}}$, between excess charge
on the ionized atom and hydrogen because this part is exactly the same as the H-H$^+$ interaction. Indeed, Uns\"old (\cite{Unsold1}) and Dalgarno \& Lewis (\cite{Dalgarno_56}, equation 16) showed that $E_p=-4/9$ for the long-range H-H$^+$ interaction. For the part of the interaction describing the interaction between the ion without the excess charge and hydrogen atom, the Uns\"old value of $-4/9$ cannot be expected to be a good approximation (Barklem \& O'Mara \cite{Barklem2bis}).  The reason that a value of $-4/9$ works well for neutrals is the fact that the separations of energy levels of the perturbed neutral atom are small compared to the separations between the ground level and the excited levels of the hydrogen atom, and thus the denominators are dominated by contributions from the H energy levels. For ions this is not the case. As a result
of the increased core charge, the energy level spacings are generally much 
larger than for neutrals. It necessary therefore to determine $E_p$ directly for each state of the ion. The appropriate value of $E_p$  can be found  via:
 \begin{eqnarray} \label{eq6}
E_p = - \frac{2\langle p^2_2 \rangle} {C_6}, 
\end{eqnarray}
where $C_6$ is  the Van der Waals constant $C_6$ averaged over all $m$ substates. The  $C_6$  coefficient is given by the standard expression  (see for example, Goodisman \cite{Goodisman_73}):
\begin{eqnarray} \label{eq7}
C_6=\frac{3}{2} \sum_{k^\prime\neq k}\sum_{l^\prime\neq l}
\frac{f^H_{kk^\prime} f^A_{ll^\prime}}
{(E_{k^\prime}^H + E_{l^\prime}^A - E^H_k - E^A_l)
(E_{k^\prime}^H - E^H_k ) (E_{l^\prime}^A - E^A_l)}
\end{eqnarray}
$f^A_{ll^\prime}$ and $f^H_{kk^\prime}$ are the dipole oscillator strengths of all transitions to the state of interest $l$ for the perturbed ion and the ground state $k$ for the neutral hydrogen atom.  $E^H$ and $E^A$  are the  energy 
eigenvalues of the hydrogen and ionised atom respectively.
More details about the calculation of $C_6$ are given in Barklem \& O'Mara (\cite{Barklem2bis}) and references therein.   The quantity  $\langle p^2_2 \rangle$ is the mean square distance between the valence electron and the perturbed  ion  core located at A (Fig. \ref{perturbed}),
\begin{eqnarray} \label{eq8}
\langle p^2_2 \rangle  = \int_{0}^{+\infty} P^2_{n^*l}(p_2) \; p^2_2 \; dp_2,
\end{eqnarray}
$P_{n^*l}$ are the the radial wavefunctions (note $P_{n^*l}(p_2)=R_{n^*l}(p_2) p_2$) of the valence electron of the 
perturbed atom (Anstee \cite{Anstee2}, Seaton \cite{Seaton_58}). $n^*$ is the effective principal quantum number corresponding to  the state $|n l  \rangle$ of the 
valence electron (Papers I, II, III). 

Using the Uns\"old approximation the expression for $V_{\textrm{eff}}$ becomes
\begin{eqnarray} \label{eq9}
\langle M_l | V_\textrm{{eff}} | M_l \rangle & = & \langle 
M_l \big\vert V\big\vert M_l \rangle - \frac{1}{E_p}(\langle M_l \big\vert 
V\big\vert M_l \rangle )^2 +  \frac{1}{E_p} \langle M_l \big\vert V^2\big\vert
 M_l \rangle \nonumber \\
&+&  \frac{1}{E_p}(\langle M_l \big\vert V_{\textrm{ind}}\big\vert M_l \rangle )^2 -  \frac{1}{E_p} \langle M_l \big\vert V^2_{\textrm{ind}}\big\vert M_l \rangle. \\
&-& \frac{9}{4}[\langle M_l \big\vert V^2_{\textrm{ind}}\big\vert M_l \rangle - (\langle M_l \big\vert V_{\textrm{ind}}\big\vert M_l \rangle )^2] \nonumber
\end{eqnarray}
$V_{\textrm{eff}}$ of equation (\ref{eq9})  is the so-called Rayleigh-Schr\"odinger-Uns\"old (RSU) potential 
(ABO Papers). For computing $V_{\textrm{eff}}$ it is  essential to determine $E_p$ in 
an independent calculation, as seen in Barklem \& O'Mara (\cite{Barklem2bis, Barklem2000}).  Thus for ionized atoms it is not possible to tabulate cross-sections as for neutral atoms (Papers I, II, III). Any calculations for depolarization and transfer of polarization involving ions must proceed line by line.

\section{Determination of depolarization and polarization transfer rates} 
Considering a collision between a perturbed  ion A  and 
hydrogen atom H (Fig. \ref{perturbed}).  Calculation of  the depolarization and transfer rates follows essentially  the steps listed below:
\begin{enumerate}
\item calculation of the required atomic wavefunctions of the system A+H 
\item determination of $E_p$ directly for each state of the ion using equation (\ref{eq6})   
\item numerical evaluation of the RSU interaction energy of the system A+H given by
equation (\ref{eq9})
\item use of these interaction potentials in the Schr\"odinger equation
describing the evolution of A+H collsional system in 
order to 
obtain the probabilities of depolarization and polarization transfer for a given impact parameter and a 
relative velocity (more details in Paper I; see also Papers II, III)
\item calculation of depolarization and polarization transfer cross-sections for each relative velocity by integration over 
impact parameters 
\item  integration of cross-sections over the Maxwell distribution of velocities 
to obtain the semi-classical depolarization and polarization transfer rates 
for a range of local temperatures of the medium 
\end{enumerate}

\section{Depolarization and polarization transfer rates for Ca$^+$-H system}
\begin{figure}[htbp] 
\begin{center}
\includegraphics[width=8 cm]{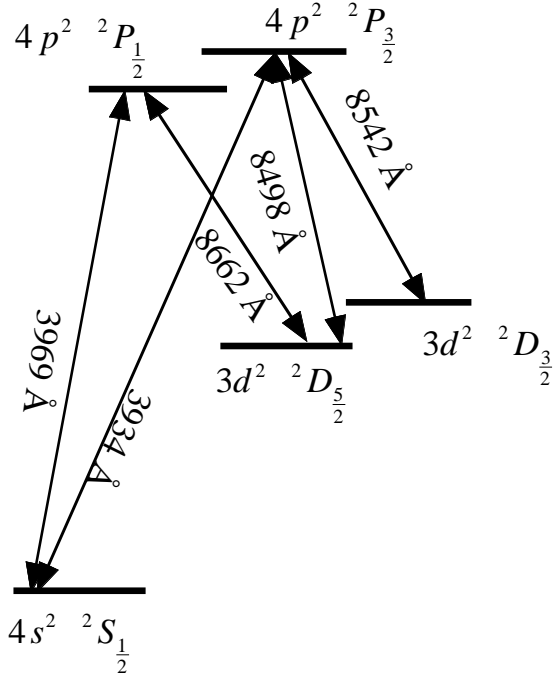}
\end{center}
\caption{Partial  Grotrian diagram of CaII showing the  levels and the spectral wavelengths in \AA $\;$ of interest in this study. Note that the level spacings in not to scale.}
\label{CaIIStructure}
\end{figure}
An important  point to emphasise is that this  semi-classical method for the calculation of depolarization and polarization transfer rates  is not specific for a given perturbed atom or ion. This method can be applied for any perturbed ion,
but we must calculate the $E_p$ value for each case (section 3). Let us consider the case of the Ca$^+$-H system in view of its importance in astrophysics and because
it is possible to  compare with recent calculations  employing the quantum chemistry approach (Kerkeni \cite{Kerkeni_03}). The term levels associated to the  the IR triplet lines of 
 CaII (8498 $\AA$, 8542 $\AA$, and 8662 $\AA$) are $4p \; ^2P_{1/2}$, $4p \; ^2P_{3/2}$, $3d \; ^2D_{3/2}$ and $3d \; ^2D_{5/2}$ (Figure \ref{CaIIStructure}). The H and K lines occur at 3969 $\AA$ and 3933 $\AA$ (Figure \ref{CaIIStructure}); their upper states are also the upper states of the IR triplet.   Table \ref{ValuesCaII} lists, for the states of interest in this work,  $\langle p^2_2 \rangle$, $C_6$ and the  corresponding $E_{p}$ calculated via equation (\ref{eq6}) (see Barklem \& O'Mara \cite{Barklem2bis}).
\begin{table}
\begin{center}
\begin{tabular}{|l|c|c|c|c|r||}
\hline
State  & $\langle p^2_2 \rangle$ (a.u.) & $C_6$  (a.u.)& $E_p$ (a.u.)\\
\hline
3$d$ &7.54 & 12.2 & -1.236 \\
\hline
4$p$ & 22.25 &81.8&-0.544 \\
\hline
\end{tabular}
\end{center}
\caption{Average energy $E_p$ for the interaction of CaII  3$d$ and 4$p$  
states with hydrogen in its ground state together with $\langle p^2_2 \rangle$ and $C_6$ values.}
\label{ValuesCaII}
\end{table} 
\subsection{Depolarization rates}
\begin{figure}[htbp]
\begin{center}
\includegraphics[width=8 cm]{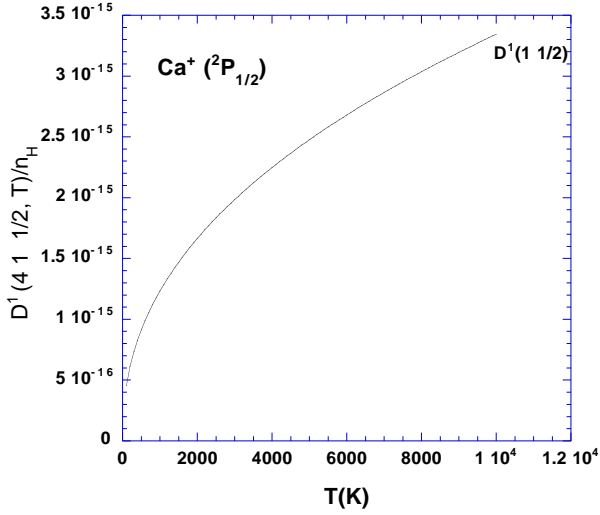}
\end{center}
\caption{Destruction rate of orientation per unit H-atom density for the CaII ion, $D^1(4 \; 1 \; 1/2, \; T)/n_H$, as a function of temperature $T$. $D^1(4 \; 1 \; 1/2, \; T)$/$n_H$  is given in $\textrm{rad.} \;  \textrm{m}^3 \; \textrm{s}^{-1}$.}
\label{depolarizationP12CaII}
\end{figure}

\begin{figure}[htbp]
\begin{center}
\includegraphics[width=8 cm]{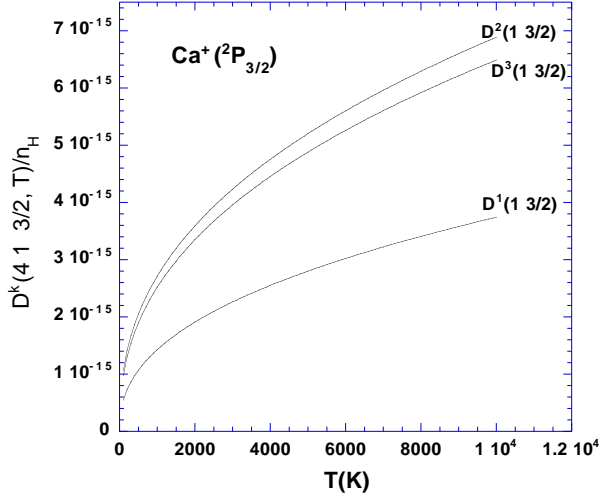}
\end{center}
\caption{Depolarization rates  per unit H-atom density for the CaII ion,  $D^k(4 \; 1 \; 3/2, \; T)/n_H$ ($k$=1, 2, and 3), as a function of temperature $T$. $D^k(4 \; 1 \; 3/2, \; T)$/$n_H$  are given in  $\textrm{rad.} \;  \textrm{m}^3 \; \textrm{s}^{-1}$.}
\label{depolarizationP32CaII}
\end{figure}

\begin{figure}[htbp]
\begin{center}
\includegraphics[width=8 cm]{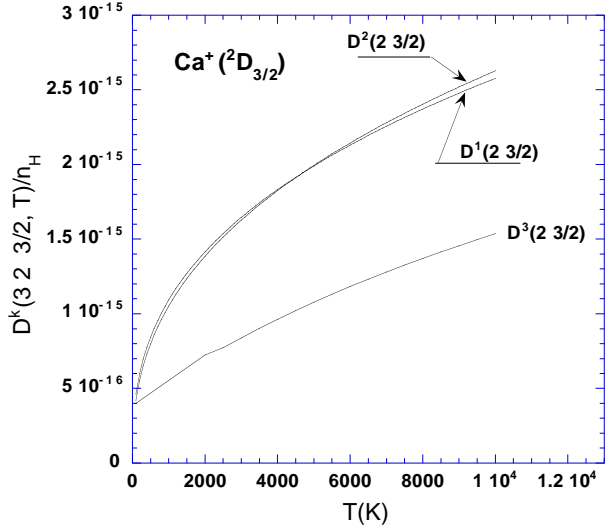}
\end{center}
\caption{Depolarization rates  per unit H-atom density for the CaII ion,  $D^k(3 \; 2 \; 3/2, \; T)/n_H$ ($k$=1, 2, and 3), as a function of temperature $T$. $D^k(3 \; 2 \; 3/2, \; T)$/$n_H$  are given in  $\textrm{rad.} \;  \textrm{m}^3 \; \textrm{s}^{-1}$.}
\label{depolarizationD32}
\end{figure}

\begin{figure}[htbp]
\begin{center}
\includegraphics[width=8 cm]{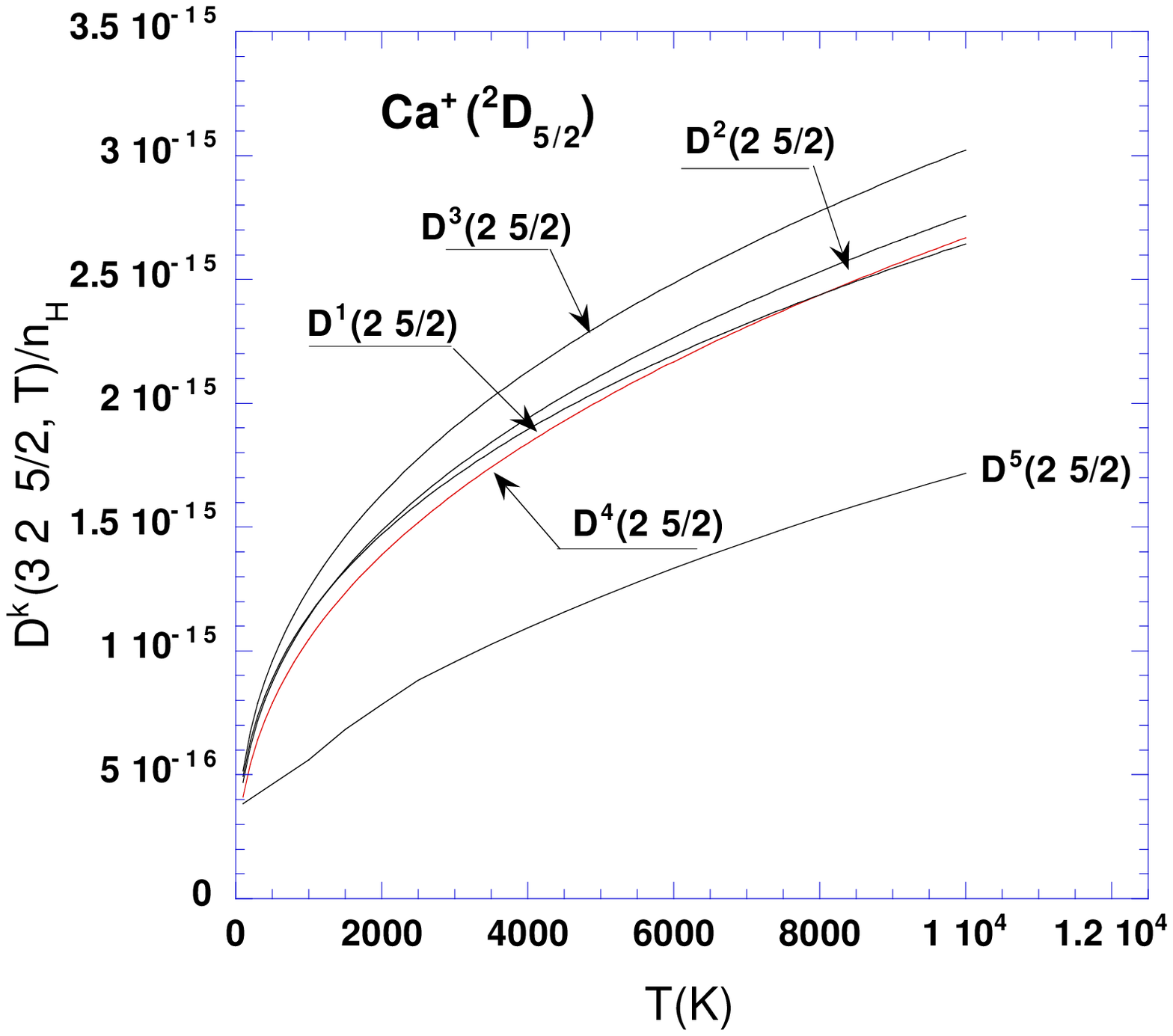}
\end{center}
\caption{Depolarization rates  per unit H-atom density for the CaII ion,  $D^k(3 \; 2 \; 5/2, \; T)/n_H$ ($k$=1, 2, 3, 4, and 5), as a function of temperature $T$. $D^k(3 \; 2 \; 5/2, \; T)$/$n_H$  are given in  $\textrm{rad.} \;  \textrm{m}^3 \; \textrm{s}^{-1}$.}
\label{depolarizationD52}
\end{figure}

The depolarization transition probability  is given by (Paper I; Sahal-Br\'echot \cite{Sahal2}):
\begin{eqnarray}  \label{eq10} 
\langle P^k(nl J, b, v) \rangle_{av}  = \frac{1}
{2J+1} \sum_{\mu,\mu'}|\langle n \;l \; J \; \mu |T|n \;l \; J \; \mu'\rangle|^2 \nonumber \\
-\sum_{\mu,\mu' , \nu,\nu'} \langle n \;l \; J \; \mu |T|n \;l \; J \; \mu'\rangle  \langle n \;l \; J \; \nu |T|n \;l \; J \; \nu'\rangle^* \quad \quad \quad \\
\times \; \sum_{\chi} (-1)^{2J+k+\mu-\mu'}
\left(
\begin{array}{ccc} 
J & J &  k  \\
-\nu' &  \mu' & \chi  
\end{array}
\right)
\left(
\begin{array}{ccc} 
J & J &  k \\ 
\nu &  -\mu & -\chi  
\end{array}
\right) \nonumber
\end{eqnarray}
Owing to the 
selection rules for the $3j$-coefficients, the summation over $\chi$ is reduced to a single term, since $\chi = -(\mu'-\nu')=-(\mu-\nu)$.
Integration over   the impact-parameter $b$ and the velocity distribution for a temperature $T$ of the medium can be performed to obtain the depolarization rate which is given by: 
\begin{eqnarray} \label{eq11}
D^k(n \;l\;  J, \; T) \simeq n_H \int_{0}^{\infty}v f(v) dv  
 \Bigg( \pi b_0^2  
+ 2 \pi \int_{b_0}^{\infty}  \langle  P^k(n \;l \; J, b, v)  \rangle_{av} 
 \;b\; db \Bigg) 
\end{eqnarray}
where $b_0$ is the cutoff impact-parameter and we use $b_0=3 a_0$ as 
in Anstee \& O'Mara (\cite{Anstee1}). 
The excited state $4p$ $^2P_{1/2}$ corresponds to total angular momentum $J=1/2$, the only non-zero  depolarization rate  is   $D^1(4 \; 1 \; 1/2, \; T)$. Figure \ref{depolarizationP12CaII} shows  $D^1(4 \; 1 \; 1/2, \; T)$  as a function of the local temperature $T$. 
The non-zero depolarization rates for the $4p \; ^2P_{3/2}$ state are $D^1(4 \; 1 \; 3/2,  \; T)$, $D^2(4 \;  1 \; 3/2, \; T)$ and $D^3(4 \; 1 \; 3/2, \; T)$, and these  rates are displayed in Figure \ref{depolarizationP32CaII}. 
The non-zero depolarization rates associated to  the $3d$ $^2D_{3/2}$ and $3d$ $^2D_{5/2}$ states are $D^1(3 \; 2 \; 3/2, \; T)$, $D^2(3 \; 2 \; 3/2, \; T)$, $D^3(3 \; 2 \; 3/2, \; T)$ for $3d$ $^2D_{3/2}$ and $D^1(3 \;  2 \; 5/2, \; T)$, $D^2(3 \;  2 \; 5/2, \; T)$, $D^3(3 \;  2 \; 5/2, \; T)$, $D^4(3 \;  2 \; 5/2, \; T)$ and $D^5(3 \;  2 \; 5/2, \; T)$ for $3d$ $^2D_{5/2}$ (see Figures \ref{depolarizationD32} and \ref{depolarizationD52}). 

\vspace{1cm}
All of the rates for the $4p$ $^2P_{1/2}$, $4p$ $^2P_{3/2}$, $3d$ $^2D_{3/2}$ and $3d$ $^2D_{5/2}$ states of CaII are found to increase with temperature
 in the range $100 \leq$ $T \leq $ 10000 K. As for neutral atoms, a functional form $\displaystyle D(T) = B T ^{(1-\lambda)/2}$ can usually be accurately fit to these depolarization rates, where $\lambda$ is the so-called velocity exponent  (Papers I, II, III). We find the following analytical expressions for the depolarization rates for $100 \le T \le 10000$ K (except for  $D^3(3 \;  2 \; 5/2, \; T)$
and $D^5(3 \; 2 \; 5/2, \; T)$ which are given for $2500 \le T \le 10000$ K):
\begin{itemize}
\item {\bf CaII($4p$ $^2P_{1/2}$)-H$(1s)$}:
\begin{eqnarray}  
 D^1(4 \; 1 \; 1/2, \; T) &=& 2.4767 \times 10^{-15} \; n_H (\frac{T}{5000})^{0.433} \; (\textrm{rad.}   \textrm{m}^3  \textrm{s}^{-1}).
\end{eqnarray}
\item {\bf CaII($4p$ $^2P_{3/2}$)-H$(1s)$}:
\begin{eqnarray} \label{eq12}
D^1(4 \; 1 \; 3/2, \; T) & = & 2.7993 \times 10^{-15} \; n_H \; (\frac{T}{5000})^{0.418} \; (\textrm{rad.}   \textrm{m}^3  \textrm{s}^{-1}) \nonumber \\
D^2(4 \; 1 \; 3/2, \; T) & = & 5.2034 \times 10^{-15} \; n_H \;  (\frac{T}{5000})^{0.405} \; (\textrm{rad.}   \textrm{m}^3  \textrm{s}^{-1}) \\
D^3(4 \; 1 \; 3/2, \; T) & = & 4.8807  \times 10^{-15}\; n_H \; (\frac{T}{5000})^{0.411}\; (\textrm{rad.}   \textrm{m}^3  \textrm{s}^{-1}) \nonumber
\end{eqnarray}
\item {\bf CaII($3d$ $^2D_{3/2}$)-H$(1s)$}:
\begin{eqnarray} \label{eq13}
D^1(3 \; 2 \; 3/2, \; T) & = & 1.9904 \times 10^{-15}\; n_H \; (\frac{T}{5000})^{0.373} \; (\textrm{rad.}   \textrm{m}^3  \textrm{s}^{-1}) \nonumber \\
D^2(3 \; 2 \; 3/2, \; T) & = & 1.9943 \times 10^{-15} \; n_H \; (\frac{T}{5000})^{0.398} \; (\textrm{rad.}   \textrm{m}^3  \textrm{s}^{-1}) \\
D^3(3 \; 2 \; 3/2, \; T) & = & 1.0772  \times 10^{-15}\; n_H \; (\frac{T}{5000})^{0.501} \; (\textrm{rad.}   \textrm{m}^3  \textrm{s}^{-1}) \nonumber,
\end{eqnarray} 
\item {\bf CaII($3d$ $^2D_{5/2}$)-H$(1s)$}:
\begin{eqnarray} \label{eq14}
D^1(3 \;  2 \; 5/2, \; T) & = & 2.0535 \times 10^{-15}\; n_H \; (\frac{T}{5000})^{0.365} \; (\textrm{rad.}   \textrm{m}^3  \textrm{s}^{-1}) \nonumber \\
D^2(3 \; 2 \; 5/2, \; T) & = & 2.1120 \times 10^{-15} \; n_H \;  (\frac{T}{5000})^{0.384} \; (\textrm{rad.}   \textrm{m}^3  \textrm{s}^{-1}) \nonumber  \\
D^3(3 \; 2 \; 5/2, \; T) & = & 2.3170  \times 10^{-15} \; n_H \;  (\frac{T}{5000})^{0.384} \; (\textrm{rad.}   \textrm{m}^3  \textrm{s}^{-1}) \\
D^4(3 \; 2 \; 5/2, \; T) & = & 2.0127  \times 10^{-15} \; n_H \; (\frac{T}{5000})^{0.407} \; (\textrm{rad.}   \textrm{m}^3  \textrm{s}^{-1}) \nonumber \\
D^5(3 \;  2 \; 5/2,  \; T) & = & 1.2187  \times 10^{-15} \; n_H \; (\frac{T}{5000})^{0.486} \; (\textrm{rad.}   \textrm{m}^3  \textrm{s}^{-1}) \nonumber.
\end{eqnarray} 
\end{itemize}

\subsection{Polarization  transfer rates}
The collisional transfer transition probability is given by (Paper II; Sahal-Br\'echot \cite{Sahal2}):
\begin{eqnarray} \label{eq15}
\langle P^k(n l J \to n l J', b, v)  \rangle_{av}& =& 
\sum_{\mu,\mu' , \nu,\nu'} \langle n l \; J \; \mu |T|n l \; J' \; \mu'\rangle  \langle n l \; J \; \nu |T|n l \; J' \; \nu'\rangle^* \\  && \sum_{\chi} (-1)^{J-J'+\mu-\mu'} 
\left(
\begin{array}{ccc} 
J & J &  k  \nonumber  \\
\nu &  -\mu & \chi  
\end{array}
\right) \left(
\begin{array}{ccc} 
J' & J' &  k \\ 
\nu' &  -\mu' & \chi  
\end{array}
\right).
\end{eqnarray}
As in equation (\ref{eq11}), the   polarization  transfer rates $\displaystyle D^k(n l  J \to n l J', \; T)$   follow from  integration over the 
impact parameters and the velocities with a Maxwellian distribution.

\begin{figure}[htbp]
\begin{center}
\includegraphics[width=8 cm]{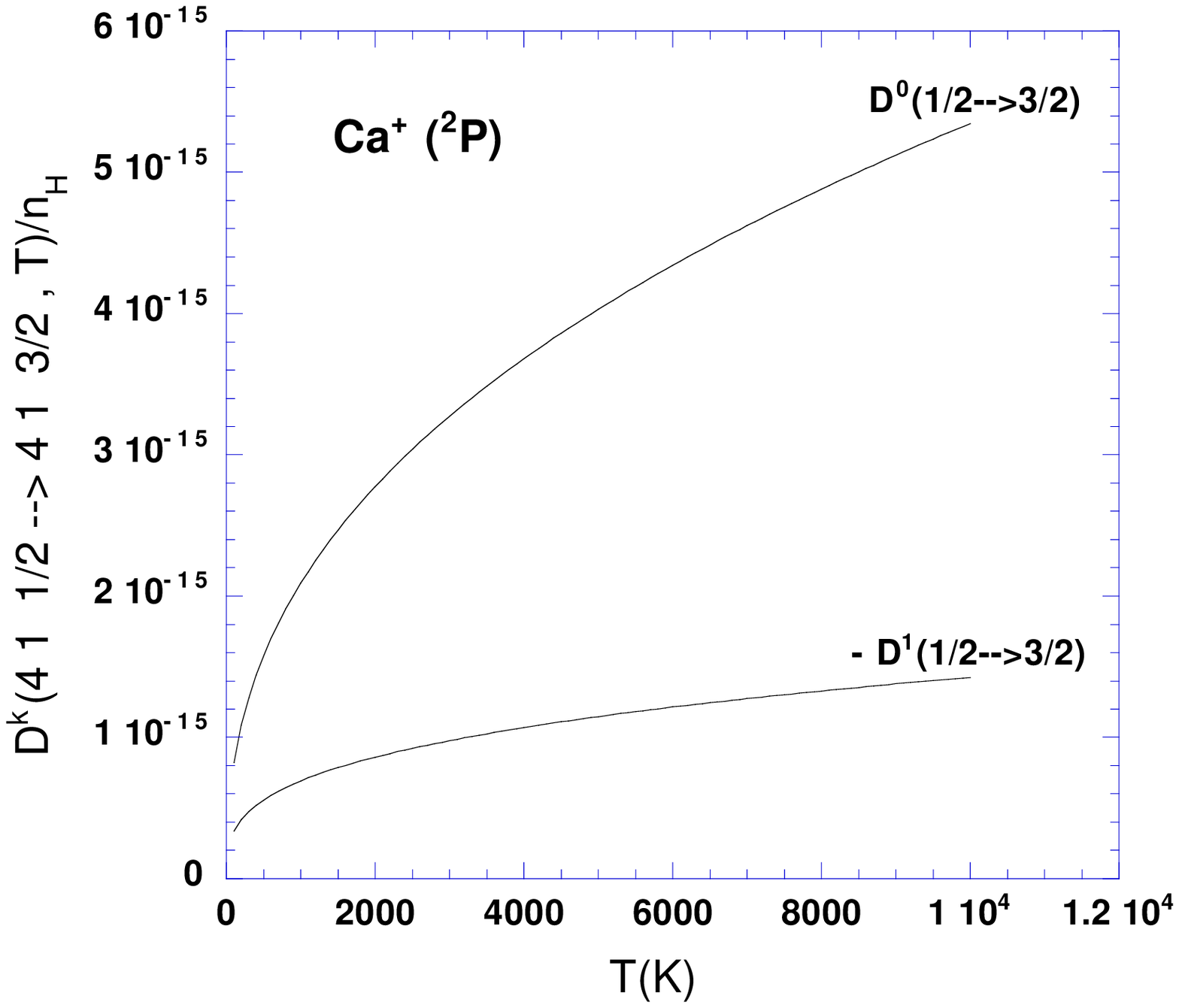}
\end{center}
\caption{Population and orientation transfer rates ($k$=0 and $k$=1), per unit H-atom density, as a function of temperature $T$. The rates are given in  $\textrm{rad.} \;  \textrm{m}^3 \; \textrm{s}^{-1}$.}
\label{depolarizationTransPCaII}
\end{figure}

\begin{figure}[htbp]
\begin{center}
\includegraphics[width=8 cm]{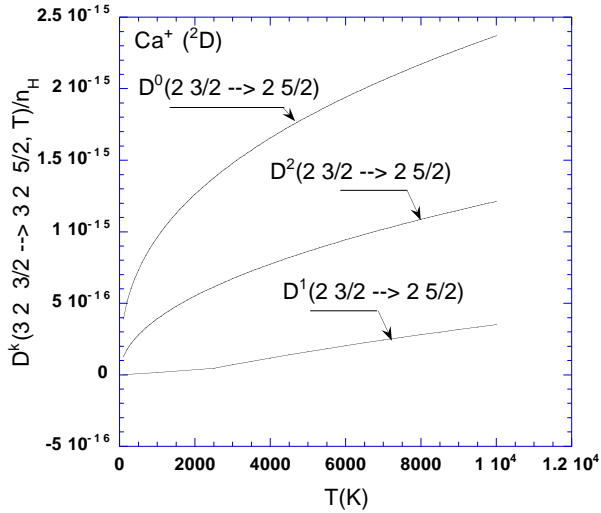}
\end{center}
\caption{Polarization transfer rates per unit H-atom density, $D^k(3 \;  2 \; 3/2 \to 3 \; 2 \; 5/2,  \; T)/n_H$, as a function of temperature $T$. The rates are given in  $\textrm{rad.} \;  \textrm{m}^3 \; \textrm{s}^{-1}$.}
\label{depolarizationtransferD}
\end{figure}

Inelastic collisions with neutral hydrogen which leave the radiating atom in 
a final state $n'l'$  different from the initial one $nl$ are neglected. 
Only the  polarization transfer rates inside the subspace $nl$ are taken into account. Our transfer rates between the levels $4p$ $^2P_{1/2}$ $\to$ $4p$ $^2P_{3/2}$, $3d$ $^2D_{3/2}$ $\to$ $3d$ $^2D_{5/2}$  ($D^k(4 \; 1 \; 1/2 \to 4 \; 1 \; 3/2, \; T)$ and $D^k1(4 \; 1 \; 1/2 \to 4 \; 1 \; 3/2, \; T)$) are presented in figures 
\ref{depolarizationTransPCaII}  and \ref{depolarizationtransferD} respectively. $D^3(3 \;  2 \; 3/2 \to 3 \;  2 \; 5/2, \; T)$  did not obey a power law of the form $B$ $T^{(1-\lambda)/2}$. However, we can provide the analytical expressions  for the other non-zero transfer rates:
\begin{eqnarray} \label{eq16}
D^0(4 \;  1 \; 1/2 \to 4 \;  1 \; 3/2,  \; T) & = & 4.0307 \times 10^{-15} 
n_H (\frac{T}{5000})^{0.407} (\textrm{rad.}   \textrm{m}^3  \textrm{s}^{-1}) 
\nonumber \\ 
D^1(4 \;  1 \; 1/2 \to 4 \;  1 \; 3/2,  \; T) & = & - 1.1464 \times 10^{-15} n_H 
(\frac{T}{5000})^{0.314} (\textrm{rad.}   \textrm{m}^3  \textrm{s}^{-1}) 
\nonumber \\
D^0(3 \;  2 \; 3/2 \to 3 \;  2 \; 5/2,  \; T) & = &  1.8061 \times 10^{-15} n_H 
(\frac{T}{5000})^{0.392} (\textrm{rad.}   \textrm{m}^3  \textrm{s}^{-1}) \\
D^1(3 \;  2 \; 3/2 \to 3 \;  2 \; 5/2, \;  T) & = & 1.6177  \times 10^{-16}  n_H 
(\frac{T}{5000})^{1.401} (\textrm{rad.}   \textrm{m}^3  \textrm{s}^{-1}) 
\nonumber \\
D^2(3 \;  2 \; 3/2 \to 3 \;  2 \; 5/2,  \; T) & = & 8.6286 \times 10^{-16} n_H 
(\frac{T}{5000})^{0.490} (\textrm{rad.}   \textrm{m}^3  \textrm{s}^{-1}) 
\nonumber 
\end{eqnarray}
\section{Comparisons}
It is important to notice that  the depolarization rates $D^k(n \;  l \; J,  \; T)$ as defined in this work (equation \ref{eq11})  and  the relaxation rates $g^k(J)$ as defined by Kerkeni et al. (\cite{Kerkeni_03}) (equation 2, section 3.2) 
are {\it not equivalent}.  Kerkeni et al. (\cite{Kerkeni_03}) defines the depolarization cross-section (or relaxation rate $g^k(J)$) as the sum of two terms: the term responsible exclusively for the depolarization of the level $(nlJ)$ and the term corresponding to the fine structure transfer between the levels $(nlJ)$ $\to$ $(nlJ')$. In our definition, $D^k(n \;  l \; J,  \; T)$ is only the depolarization of the level $(nlJ)$, the fine structure transfer is not included. We calculate separately the rates associated to fine structure transfer which are $k$-independent and proportional to the population transfer rates $D^0(n \;  l \; J \to n \;  l \; J', \; T)$ (equation 4 of Paper II). 

\begin{figure}[htbp]
\begin{center}
\includegraphics[width=8 cm]{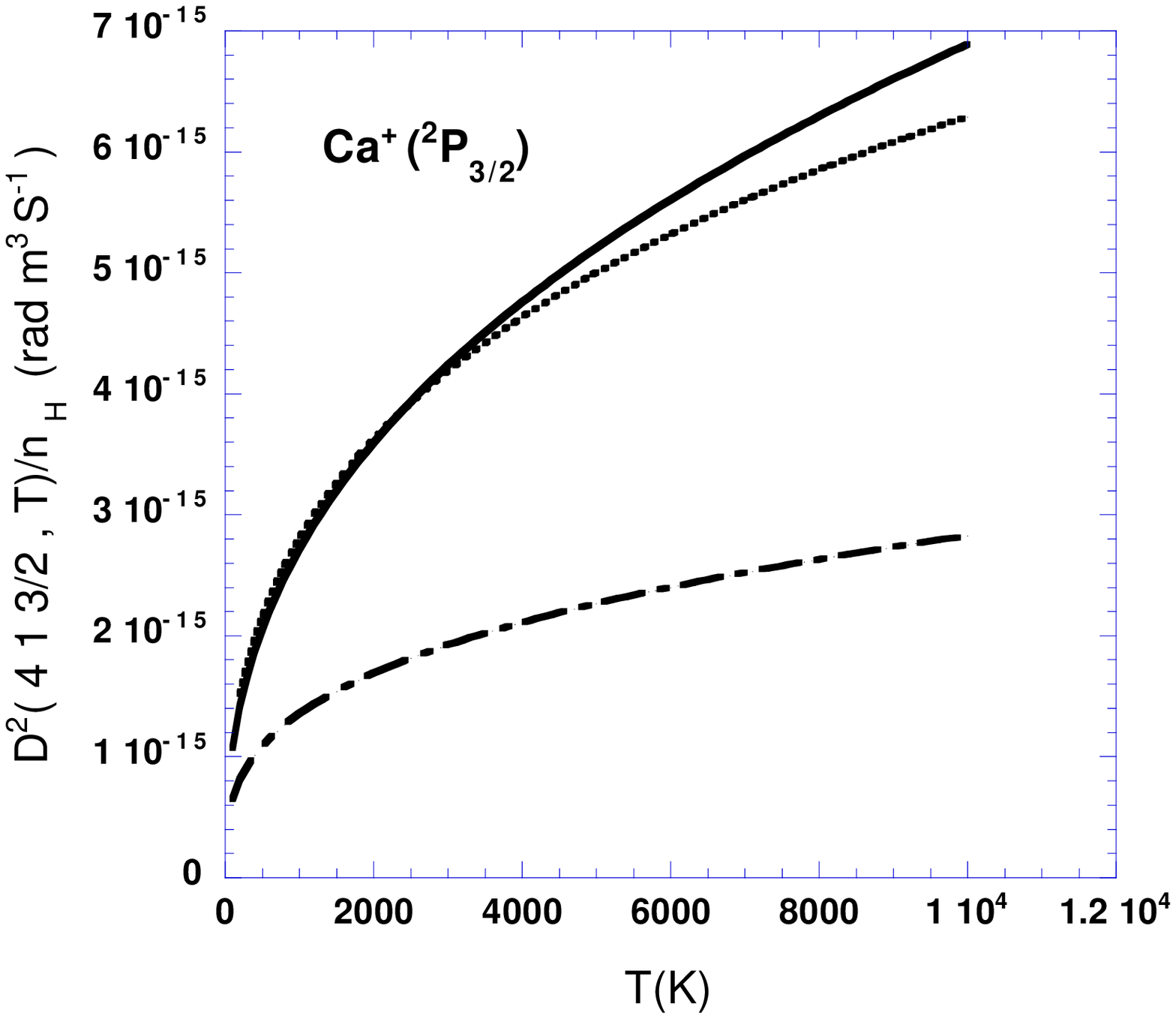}
\includegraphics[width=8 cm]{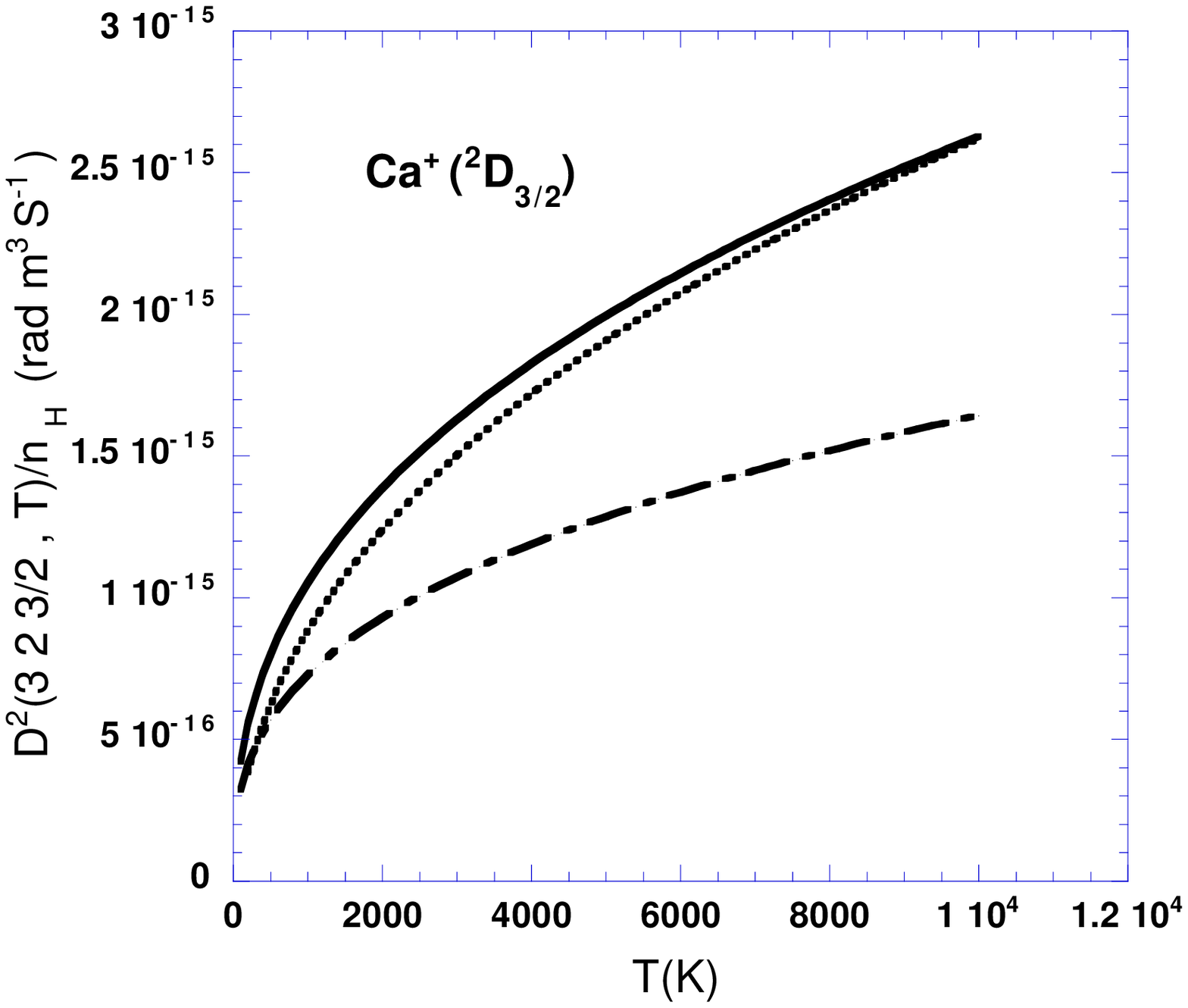}
\includegraphics[width=8 cm]{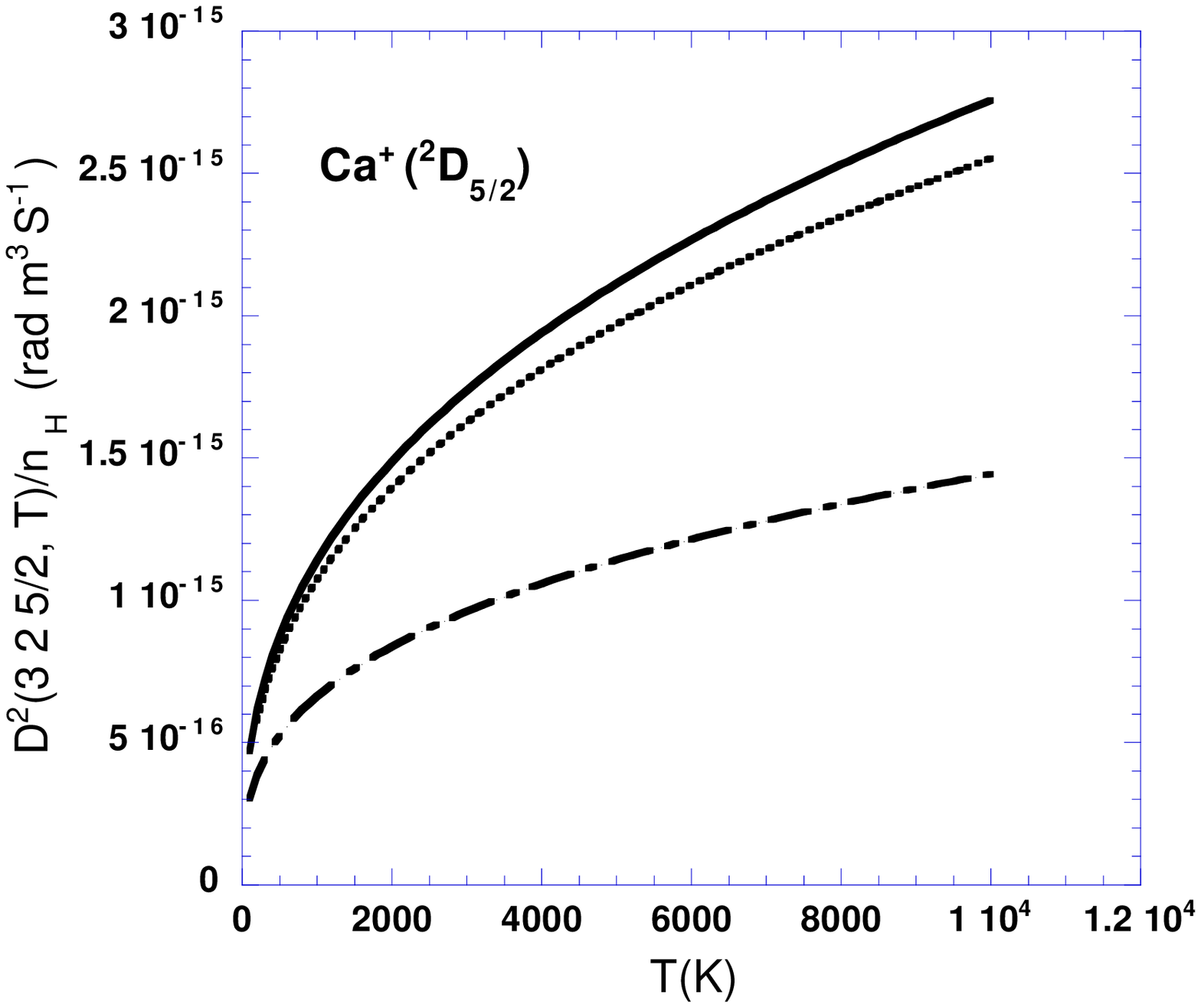}
\end{center}
\caption{Depolarization rates for $k=2$ as a function of temperature. Full lines: our results; dotted lines: quantum chemistry calculations (Kerkeni \cite{Kerkeni_03}); dot-dashed lines  Van der Waals approximation.}
\label{depolarizationk2D52}
\end{figure}

For example, in 
accordance with our definition $D^0(n \;  l \; J,  \; T) \equiv 0$ but  the 
relaxation rates  $g^0(J)$ from Kerkeni et al. (\cite{Kerkeni_03}) are not necessarily zero. In 
order to compare to  the Kerkeni et al. (\cite{Kerkeni_03}) results  it is essential to 
substract the part of $g^k(J)$ associated to fine structure transfer ($g(J \to J')$ in   Kerkeni et al. \cite{Kerkeni_03}).  This  it is nothing more than a difference in  
definitions. Nevertheless, this difference should be taken into account when writing 
the SEE. 
We now compare  our alignment depolarization rates with the quantum chemistry depolarization rates (Kerkeni et al. \cite{Kerkeni_03}) and the alignment depolarization rates obtained by replacing the RSU potential with the Van der Waals potential, $\displaystyle V = -\frac{C_6}{R^6}$, where $C_6$ is taken from Table~1. Note that the latter rates differ from the usual Van der Waals formula for the rates, in that while they employ the van der Waals potential the collision dynamics are treated using our theory, and the $C_6$ value is accurately determined (whereas typically the approximation $C_6=\alpha_H \langle p_2^2 \rangle$ is employed where $\alpha_H$ is the polarisability of hydrogen).  
In Figure  \ref{depolarizationk2D52}  we show our alignment depolarization rates with quantum chemistry depolarization rates (Kerkeni et al. \cite{Kerkeni_03}) and the improved  Van der Waals rates. We  display  only the $k=2$ case which is related to the linear 
depolarization (alignment). Reference to Figure \ref{depolarizationk2D52}  
shows that the Van der Waals potential significantly underestimates the depolarization 
cross section. Our results show rather good agreement with quantum chemistry calculations.  
 In concrete terms,  the percentage  errors  at $T$=5000 K with respect to quantum chemistry depolarization rates are 3.9 $\%$ for  $D^2(4 \;  1 \; 3/2, \; T)/n_H$; 4.6  $\%$ for $D^2(3 \;  2 \; 3/2, \; T)/n_H$ and 7.3 $\%$ for $D^2(3 \;  2 \; 5/2, \; T)/n_H$. The  errors for the other depolarization and transfer  rates  are similar to the errors for the destruction rates of alignment (in general, less than 15 $\%$).  This similarity is expected since all  
rates orginate from the same the collisional processes.   


\section{Discussion}
\subsection{Dependence of depolarization  cross-sections on interatomic separations}
\begin{figure}[htbp]
\begin{center}
\includegraphics[width=8 cm]{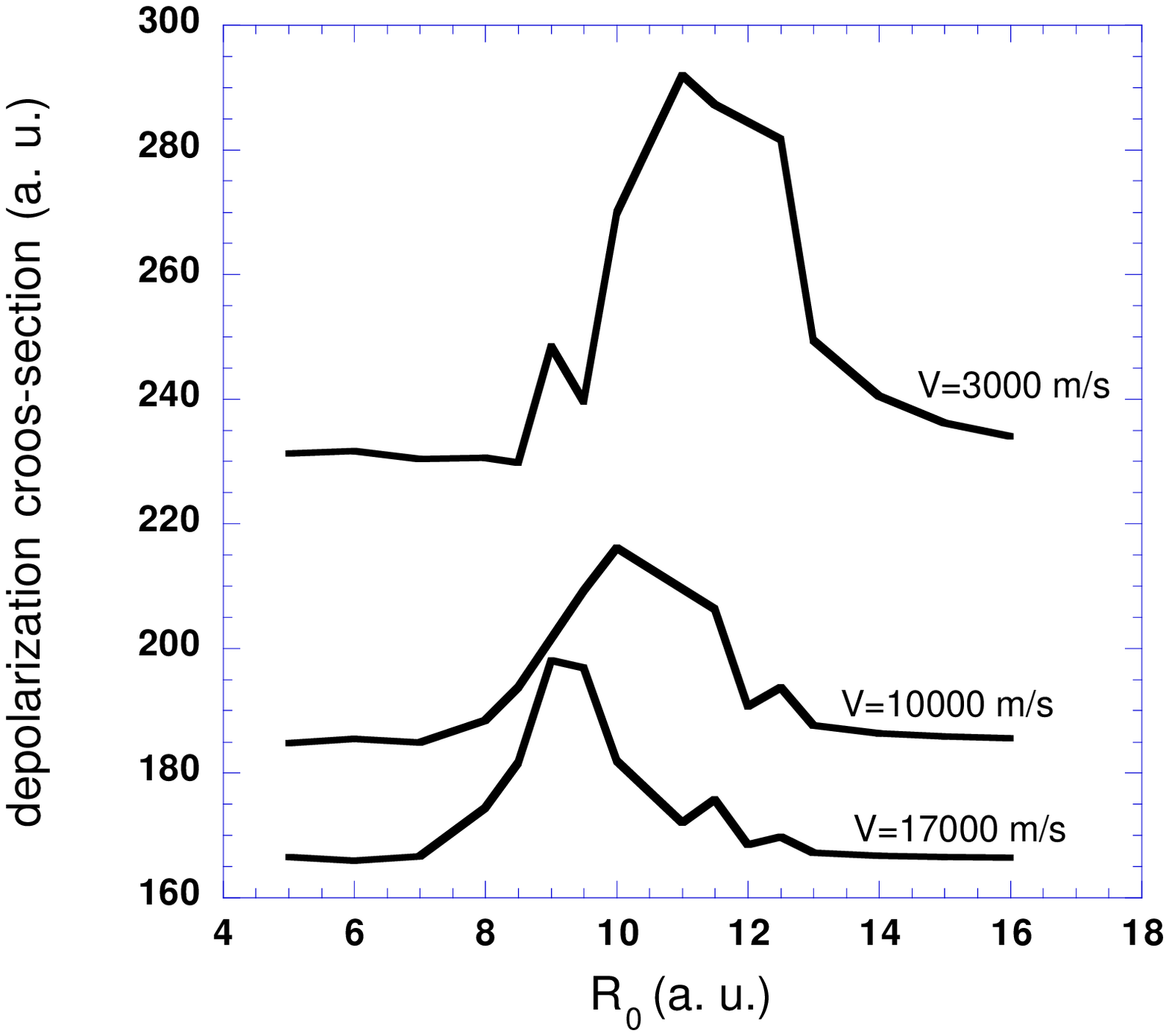}
\end{center}
\caption{Depolarization cross section enhancement due to a Gaussian local 
perturbation of the potential. Cross sections are calculated for  
the relative velocities: $3 \; \textrm{km} \; \textrm{s}^{-1}$, $ 10 \; \textrm{km} \; \textrm{s}^{-1}$ and $ 17 \; \textrm{km} \; \textrm{s}^{-1}$.}
\label{depolarizationlumpCaII}
\end{figure}
To assess the sensitivity of the calculations to the accuracy of the potentials at various separations we make calculations with potentials where we have introduced local perturbations. 
The interaction potential $(1s$, $4p\sigma)$  is multiplied by a  Gaussian magnification factor of the form  (Anstee \& O'Mara  \cite{Anstee1}): 
\begin{eqnarray} \label{eq17}
G(R)=1+\exp(-(R-R_0)^2).
\end{eqnarray} 
$R_0$ is the position where the variation of $V_{\textrm{eff}}$ reaches its  
maximum value (the interaction potential $(1s$, $4p\sigma)$ doubles). Figure \ref{depolarizationlumpCaII} shows the   depolarization  cross-section 
calculated with varying $R_0$.  It is clear 
that the values of $R_0$ inducing  depolarization  cross-section enhancement  confirm the fundamental result found already for neutral atoms: the interactions of importance for the depolarization cross-section (or depolarization rate)
 calculations are the intermediate-range interactions (Papers I, II, and III). The principal differences between the RSU potentials and those from quantum chemistry occur at  small interatomic separations. It is for this reason that  we 
obtain rather good agreement with quantum chemistry calculations.   The Van der Waals interaction potentials are inaccurate at the intermediate region, and this explains why these potentials underestimate the depolarization cross 
sections. 
\subsection{Dependence of depolarization  rates on $E_p$}
\begin{figure}[htbp]
\begin{center}
\includegraphics[width=8 cm]{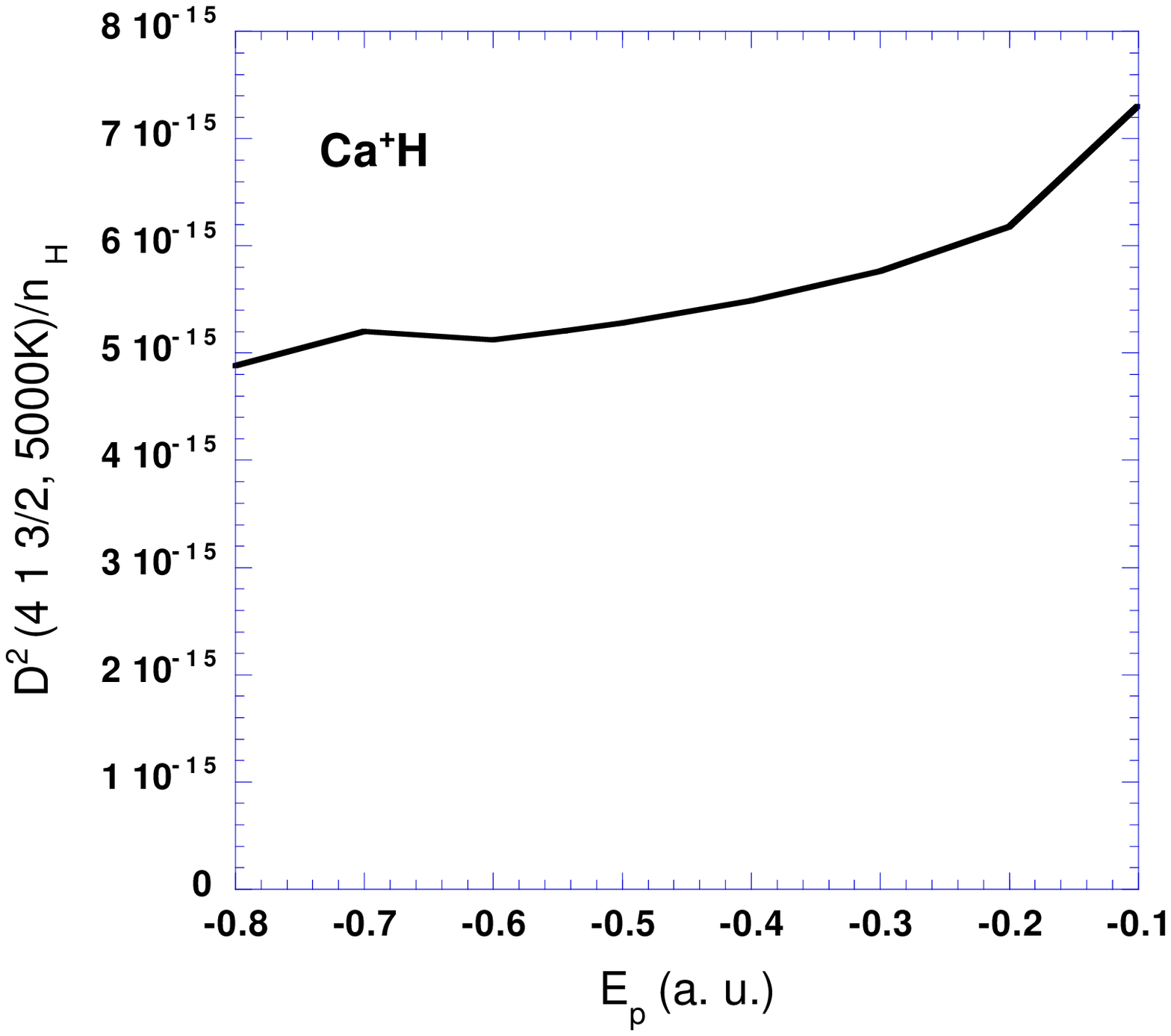}
\end{center}
\caption{Dependence on $E_p$ of the destruction rate of alignment $D^2(4 \; 1 \; 3/2,$ 5000K)/$n_H$.}
\label{depol_k2_CaII_Ep}
\end{figure}
The depolarization and transfer rates for the $4p$ and $3d$ states are calculated for $E_p=-0.544$ and $-1.236$ respectively. As a check on the sensivity of our results  to the precision of the calculations of $E_p$, we have calculated the destruction rate of alignment $D^2(4 \;  1 \; 3/2, \; 5000 \textrm{K})$ by varying  $E_p$ in equation  (\ref{eq9}).  Note that when $E_p$ decreases, the interaction potential decreases and so  $D^2(4 \;  1 \; 3/2, \; 5000 \textrm{K})$ also decreases (Figure \ref{depol_k2_CaII_Ep}). The depolarization rate shows only an extremly weak variation with $E_p$.   An $|E_p|$ variation of 25 $\%$, with respect to the value $E_p=-0.544$, corresponds to a change of less than 5 $\%$  in the calculated depolarization rates.   It should not concluded that this is a general property of the depolarization rates for all states of all ionised atoms.  Reference to the Figure \ref{depol_k2_CaII_Ep} shows a rather strong dependence of the depolarization rates on $E_p$ for $|E_p|$ $\le$ 0.25. We expect, however, that the value of $|E_p|$ is usually greater than 0.25 and the depolarization rates will not be greatly affected by possible error in the value of $E_p$ (see Barklem \& O'Mara \cite{Barklem2bis, Barklem2000}).

\section{Application for Sr II 4078 \AA $\;$   line}
\begin{figure}[htbp]
\begin{center}
\includegraphics[width=8 cm]{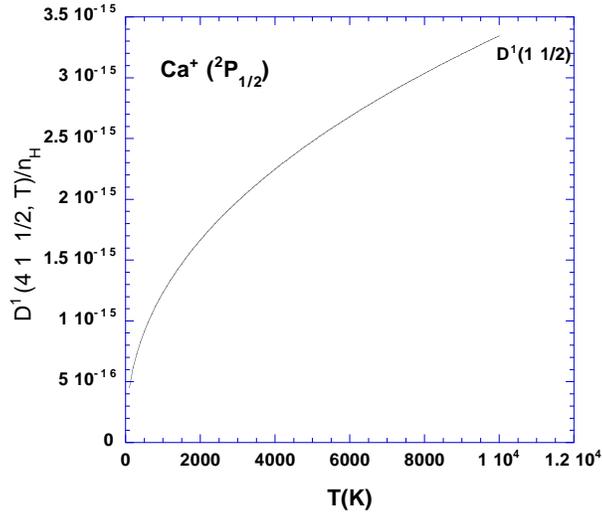}
\end{center}
\caption{Destruction rate of orientation  per unit H-atom density for the SrII ion,  $D^1(5 \; 1 \; 1/2, \; T)/n_H$, as a function of the temperature of the medium, $T$. $D^1(5 \; 1 \; 1/2, \; T)$/$n_H$  is given in $\textrm{rad.} \;  \textrm{m}^3 \; \textrm{s}^{-1}$.}
\label{depolarizationP12SrII}
\end{figure}

\begin{figure}[htbp]
\begin{center}
\includegraphics[width=8 cm]{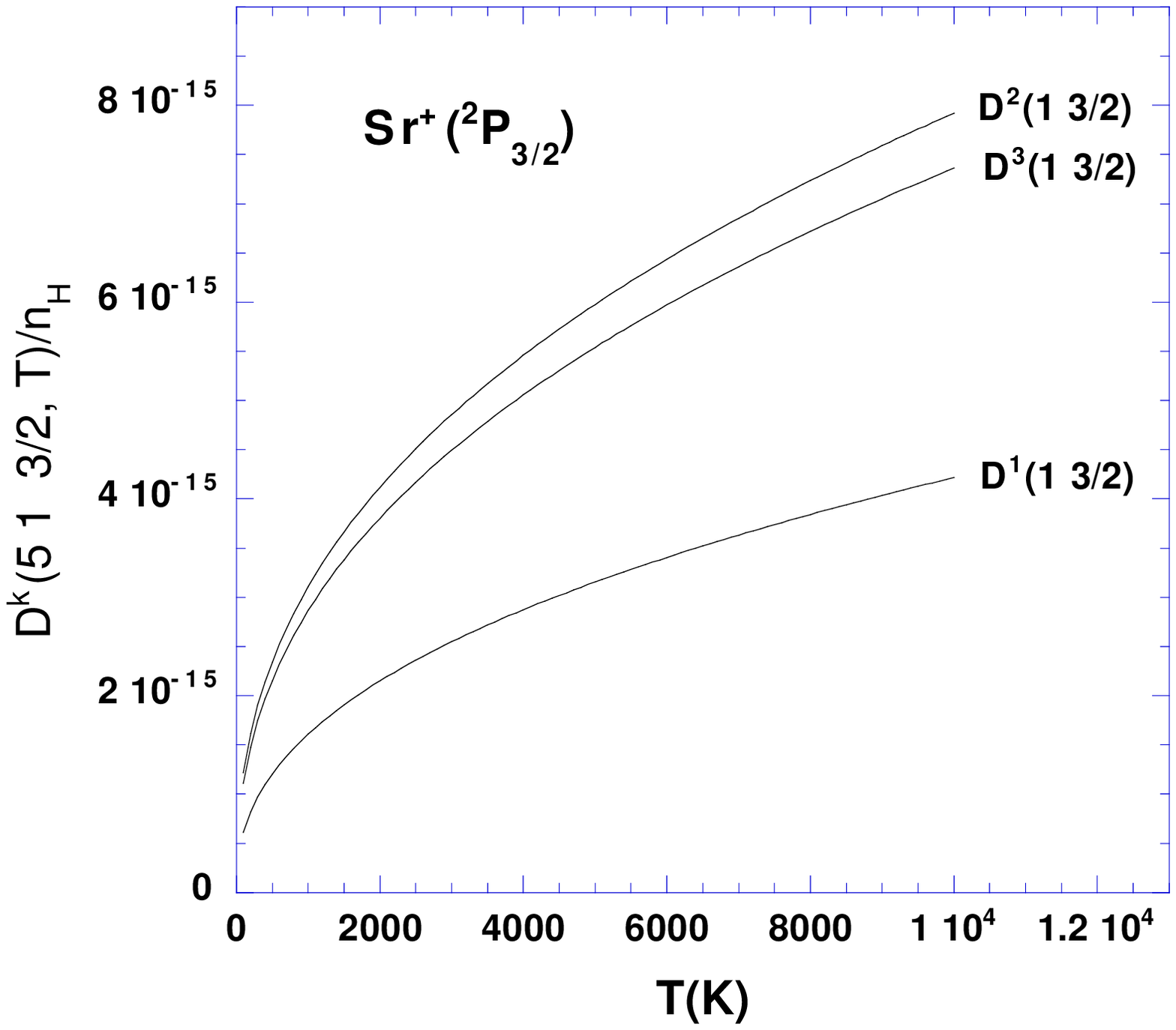}
\end{center}
\caption{Depolarization rates,   $D^k(5 \; 1 \; 3/2, \; T)/n_H$ ($k$=1, 2,  and 3), as a function of temperature $T$. $D^k(5 \; 1 \; 3/2, \; T)$/$n_H$  are given in  $\textrm{rad.} \;  \textrm{m}^3 \; \textrm{s}^{-1}$.}
\label{depolarizationP32SrII}
\end{figure}

\begin{figure}[htbp]
\begin{center}
\includegraphics[width=8 cm]{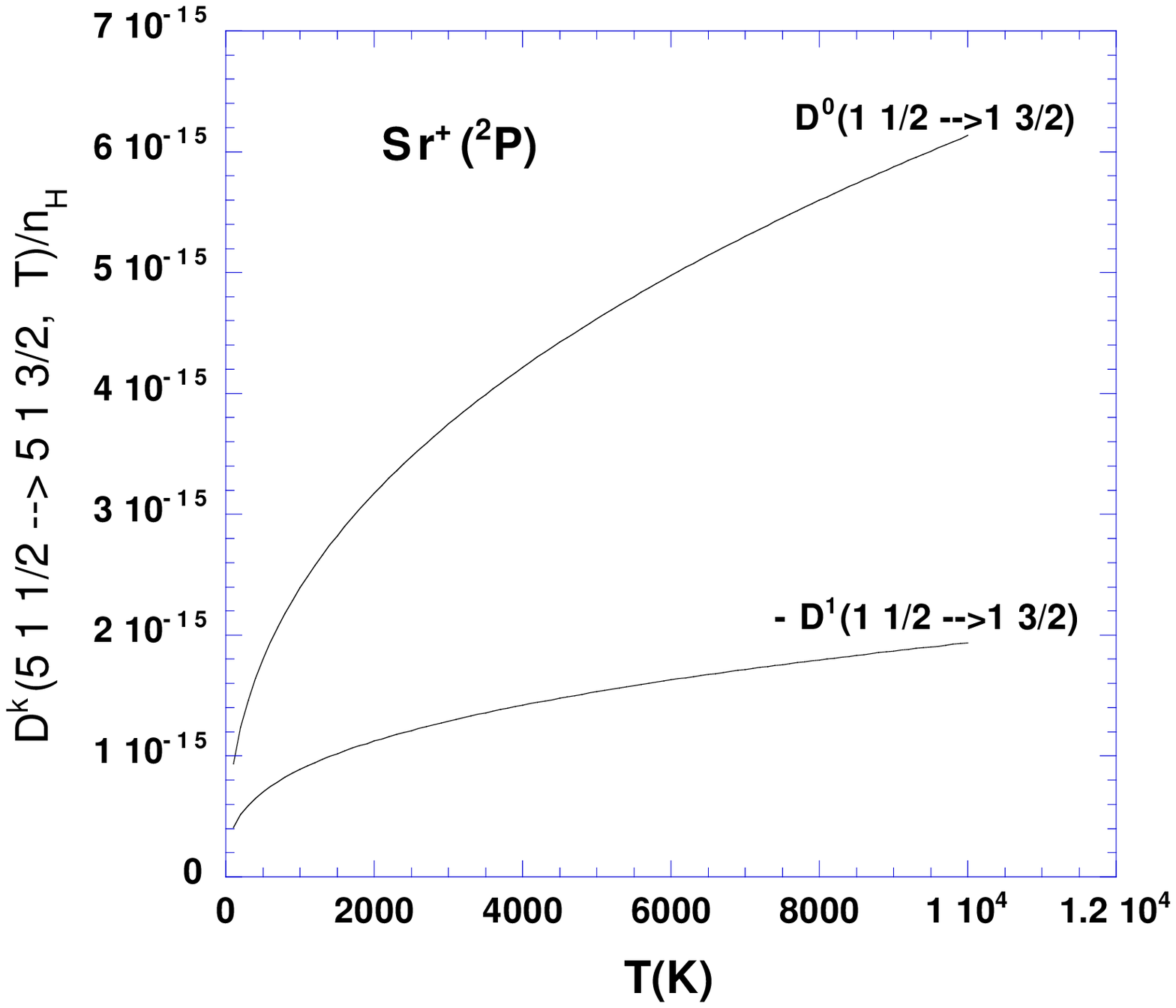}
\end{center}
\caption{Population and orientation transfer rates ($k$=0 and $k$=1) for the SrII ion, per unit H-atom density, as a function of temperature $T$. The rates are given in  $\textrm{rad.} \;  \textrm{m}^3 \; \textrm{s}^{-1}$.}
\label{depolarizationTransPSrII}
\end{figure}

The Sr II 4078 \AA $\;$   line was examined by Bianda et al. (1998),  who wrote ``{...\it The rather large uncertainty in the depolarizing collision rate
introduces a corresponding uncertainty in the field-strength scale...}''. 
These authors have used the traditional Van der Waals approach to calculate collisional 
rates.  The  4078 \AA $\;$   line is the resonance line of SrII: $5s$ $^2S$ $\to$ $5p$ $^2P$.  
We have computed the depolarization and polarization transfer rates of the levels  $5p$ $^2P_{1/2}$ and $5p$ $^2P_{3/2}$ of SrII ion.  The value of $E_p=-0.564$ for the level $5p$ $^2P$ of SrII was adopted from Barklem \& O'Mara (\cite{Barklem2000}). We applied our method to obtaining the rates which are shown in Figures \ref{depolarizationP12SrII} 
and \ref{depolarizationP32SrII}. They were found to again obey a power law $A T ^{(1-\lambda)/2}$ and are given for $100 \le T \le 10000$ K by:
\begin{itemize}
\item {\bf SrII($5p$ $^2P_{1/2}$)-H$(1s)$}:
\begin{eqnarray}  
 D^1(5 \; 1 \; 1/2, \; T)&=& 2.7196 \times 10^{-15} \; n_H \; (\frac{T}{5000})^{0.428} \; (\textrm{rad.}   \textrm{m}^3  \textrm{s}^{-1}).
\end{eqnarray}
\item {\bf SrII($5p$ $^2P_{3/2}$)-H$(1s)$}:
\begin{eqnarray}  
 D^1(5 \; 1 \; 3/2, \; T)&=& 3.1560 \times 10^{-15} \; n_H \; (\frac{T}{5000})^{0.418} \; (\textrm{rad.}   \textrm{m}^3  \textrm{s}^{-1}) \nonumber \\
 D^2(5 \; 1 \; 3/2, \; T)&=& 5.9776 \times 10^{-15} \; n_H \; (\frac{T}{5000})^{0.406} \; (\textrm{rad.}   \textrm{m}^3  \textrm{s}^{-1}) \\
 D^3(5 \; 1 \; 3/2, \; T)&=& 5.5413 \times 10^{-15} \; n_H \; (\frac{T}{5000})^{0.410} \; (\textrm{rad.}   \textrm{m}^3  \textrm{s}^{-1}). \nonumber 
\end{eqnarray}
\end{itemize}
Between the term levels $5p$ $^2P_{1/2}$ and $5p$ $^2P_{3/2}$ there are only two non-zero polarization transfer rates which are given in Figure \ref{depolarizationTransPSrII}. The analytical expressions  for these  rates for $100 \le T \le 10000$ K are:
\begin{eqnarray} \label{eq18}
D^0(5 \;  1 \; 1/2 \to 5 \;  1 \; 3/2,  T) & = & 4.6184 \times 10^{-15} \;
n_H \; (\frac{T}{5000})^{0.409} \; (\textrm{rad.}   \textrm{m}^3  \textrm{s}^{-1}) 
 \\ 
D^1(5 \;  1 \; 1/2 \to 5 \;  1 \; 3/2,  T) & = & - 1.5309 \times 10^{-15} \; n_H \;
(\frac{T}{5000})^{0.338} \; (\textrm{rad.}   \textrm{m}^3  \textrm{s}^{-1}) \nonumber
\end{eqnarray}

\section{Conclusion}
We have adapted our semiclassical method of calculation of collisional depolarization of spectral lines of neutral atoms by atomic hydrogen to allow it to be used for singly ionised atoms. Comparison with recent quantum chemistry calculations for CaII indicates and error at $T$=5000 K of less than 5 $\%$. This is an encouraging result which supports the validity of our semi-classical approach. Using this method we should be able to calculate depolarization rates of the levels involved in 
transitions of heavy
ionised atoms like  SrII, Ti II, Ce II, Fe II, Cr II, BaII...  
Calculations must proceed line by line because a suitable $E_p$ value needs to be determined for each relevant state of the given ion.  We have applied the method to calculate    
depolarization and polarization transfer rates  for the  SrII  $5p$ $^2P$ 
state. These calculations should allow a more accurate theoretical interpretation of the observed linear polarization of  the SrII 4078 \AA $\;$   line.

\end{document}